\documentclass[prd,twocolumn,preprintnumbers,superscriptaddress]{revtex4-2}
\usepackage{amsmath}
\usepackage{color}
\usepackage{ulem}
\usepackage{graphicx}
\usepackage{hyperref}
\usepackage{blindtext}
\hypersetup{colorlinks, citecolor=blue, linkcolor=blue, urlcolor=magenta}

\usepackage[caption=false]{subfig}

\usepackage[american]{babel}
\usepackage{microtype}
 \usepackage[table,xcdraw]{xcolor}
\definecolor{darkblue}{rgb}{0.0,0.0,0.6}

\newcommand{\Mpl}{M_{\rm Pl}}
\usepackage{geometry}
\usepackage{fullpage}
\begin{document}

\title{Induced Gravitational Waves via Warm Natural Inflation }

\author{Miguel~Correa}
\email{mcorrea2@nd.edu}
\affiliation{Center for Astrophysics,
Department of Physics and Astronomy, University of Notre Dame, Notre Dame, IN 46556, USA.}
\author{Mayukh R. Gangopadhyay}
\email{mayukh$\_$ccsp@sgtuniversity.org}
\affiliation{Centre For Cosmology and Science Popularization (CCSP), SGT University, Gurugram, Delhi- NCR, Haryana- 122505, India.}
\author{Nur Jaman}
\email{nurjaman.pdf@iiserkol.ac.in }
\affiliation{Department of Physical Sciences, \\
 Indian Institute of Science Education and Research Kolkata, Mohanpur - 741 246, WB, India}
\author{Grant J. Mathews}
\email{gmathews@nd.edu}
\affiliation{Center for Astrophysics,
Department of Physics and Astronomy, University of Notre Dame, Notre Dame, IN 46556, USA.}

\begin{abstract}
We analyze the  spectrum of gravitational waves generated by the induced spectrum of tensor fluctuation during warm natural inflation. In our previous work it has been  demonstrated that an epoch of warm natural inflation can lead to cosmologically relevant dark matter production in the form of  primordial black holes.  Here we show that models which solve the dark-matter production also produce a contribution to the cosmic gravitational wave background that satisfies current constraints from pulsar timing and big bang nucleosynthesis.  More importantly, this gravitational wave background may be observable in the next generation of space-based and ground-based gravitational wave interferometers. 
\end{abstract}
\maketitle
\section{Introduction}
 In our previous paper \cite{Correa22},  we investigated the intriguing properties of a cosmic inflationary paradigm \cite{inflation,lindinflation, inflation2, starobinsky} in which the inflaton effective potential is based on the Natural potential \cite{Freese}, and where there exists a coupling between the inflaton field and matter fields such that matter is continuously produced during the inflationary epoch (Warm Inflation \cite{wi1,wii1, wi2,wi9, wi3, wi4, wi5, wi6, wi7, wi8, Bastero-Gil:2019rsp, Basak:2021cgk}). This so-called ``Warm Natural Inflation" (WNI) paradigm was first studied by \cite{wnio,  wni1} and more recently by \cite{wni2, Montefalcone:2022jfw, AlHallak:2022haa}. 
In our study \cite{Correa22} we discovered that the model remarkably satisfies several observational and theoretical constraints. Firstly, it yields a spectral index and a ratio of the tensor-to-scalar power spectra that agree with constraints from the Planck mission \cite{pl1, pl2} and BICEP/Keck \cite{keck, bkplanck}. Secondly, WNI allows for inflation to be in a sub-Planckian regime within the effective field theory framework. We found it is consistent with the previously mentioned constraints for a symmetry-breaking scale of $f=0.8$ with cubic dissipation ($\Gamma \propto T^3$). Recently, \cite{Montefalcone:2022jfw} also confirmed our results by finding it is consistent for $f_{min}=0.8$ in the parameter space. However, our most important finding is that WNI naturally leads to the significant generation of primordial black holes (PBHs).  

The possibility of black hole formation in the early Universe has been a subject of consideration for decades \cite{Zeldovich:1967lct, Hawking:1971ei, Carr:1974nx}. Furthermore, it is known that if the mass of PBHs falls within an appropriate range permitted by observations, they could potentially explain the entirety of the inferred dark matter in the Universe \cite{Chapline:1975ojl, Carr:2016drx, Carr:2020xqk, Ballesteros:2017fsr}. Ref.~\cite{arya} demonstrated that warm inflation could produce sufficient enhancement in the scalar power spectrum to give rise to PBHs.  Notably, WNI generates PBHs in significant quantities and within the correct mass range to account for a substantial fraction, if not all, of the observed dark matter content in the Universe. 

 As additional motivation, it has been pointed out \cite{Cappelluti22} that a number of observational dilemmas can be understood if there is a significant population of PBHs.  Among them, in a  PBH-$\Lambda$CDM cosmology, the PBH dark matter mini-halos can collapse earlier than those comprised of standard collisionless cold dark matter.   This allows baryons to cool and form stars and galaxies at very high redshift.  This is consistent with recent JWST observations \cite{Naidu22} of bright galaxies at very high redshift ($z \sim 13)$.  The PBHs can also collect to provide seeds for supermassive black hole formation, and thereby account for the DM-halo host-galaxy central-black-hole connection as manifested in the M$_{\rm BH}-\sigma$ relation. They may also account \cite{Cappelluti22} for the X-ray and infrared backgrounds and the early formation of the super-massive black holes powering quasars at $z > 7$ .

 However, one signature we did not consider in our previous study \cite{Correa22} was the possible spectrum of gravitational waves (GWs) associated with primordial black hole production. 
 The purpose of the present work, therefore, is to examine this constraint in the context of WNI models that can account for the cosmic dark matter content.  As we will see, WNI has the capability to generate a GW spectrum that could be detectable by several future detectors.

There are two sources of primordial GWs that one needs to consider. First are the quantum tensor perturbations generated during inflation:  these are the primary GWs.  Secondly, the classical GWs generated by the enhanced density perturbation.  These are the secondary or induced GWs. In the language of perturbation theory, at linear order, the scalar and tensor modes evolve independently.  In the second order of perturbation, the scalar and tensor modes couple together.   The scalar perturbation can then source the secondary  tensor mode and thereby produce Induced Gravitational Waves (IGWs) \cite{Ananda:2006af,Mollerach:2003nq,Baumann:2007zm, Choudhury:2013woa, Kohri:2018awv}. 

Usually,  the second order GWs are suppressed with respect to the first order by a factor of the square of the scalar spectrum \cite{Kohri:2018awv}. However, they can become significant and even exceed the first-order GWs for an enhanced  primordial scalar power spectrum, such as in the case of PBH formation \cite{Ananda:2006af,baguev:001,Alabidi:2012ex}. Indeed, it has been shown in the literature \cite{Mollerach:2003nq} that the second order tensor mode dominates over the first order if the tensor to scalar ratio is $r<10^{-6}$. The induced gravity wave production in the case of warm inflation has been studied previously in \cite{marpbh, wgw1, wgw2, Arya:2023pod}. Here, we specifically consider GW production in warm natural inflation.

This paper is structured as follows: We briefly review the WNI dynamics in section \ref{wnidynamics}. The modeling of GWs and a semi-analytical calculation of the GW spectrum is given  in section \ref{InducedGW}.  We present our findings and compare the calculated spectrum to present and future detection sensitivities in section \ref{GWcons}. We provide discussion and conclusions in Section \ref{conclusion}.

%%%%%%%%%%%%%%%%%%%%%%%%%%%%%%%%%%%%%%%%%%%%%%%%%%%%%%%%%%%%%%%%%%%%%%%%%%%%####################
\section{ Warm Inflationary Dynamics }
\label{wnidynamics}
In a  homogeneous and isotropic background, the dynamics of the inflaton field $\phi(t)$ within warm inflation are governed by the following equations:
\begin{eqnarray}   
\ddot{\phi}+3 H \left(1+  Q \right)\dot{\phi}+ V_{, \phi} =0 &&~, \label{eqn1}\\
\dot{\rho}_R+4H\rho_R = 3HQ\dot{\phi}^2~,
\label{eqn2} &&\\
3 H^2 \Mpl^2  = (\rho_{\phi}+\rho_R)~, 
\label{eqn3} &&
\end{eqnarray}
where over-dots represent derivatives with respect to cosmic time $t$,~~$V_{,\phi}\equiv {\partial V}/ {\partial \phi}$, ~$\rho_r$ is the radiation energy density, and $Q\equiv {\Gamma}/{3 H}$, where $\Gamma$ is the dissipation coefficient providing the source for the radiation bath.  The last equation is the Friedmann equation 
satisfied by the Hubble parameter $H$. 

During inflation, the potential energy dominates over both the kinetic term and the radiation energy density, i.e. 
\begin{eqnarray}
 V(\phi)\gg \Big\{\frac{1}{2}\dot \phi ^2, \rho_r \Big\}\, . 
\end{eqnarray}
Also, the inflaton field amplitude should not change too quickly ($ \ddot \phi < 3 H \dot \phi$).  Moreover, the condition of an accelerating scale factor ($\ddot a > 0$) then leads to the slow-roll condition for WI such that:  
\begin{eqnarray}
    3 H(1+ Q)\approx - V_{, \phi}\,  
\end{eqnarray}
and 
\begin{eqnarray}
    - \frac{d \ln H}{dN} =& \frac{\epsilon_\phi}{1+Q} \ll 1 ~~,\\
    -\frac{d \ln\,V_{,\phi}}{d N} =&  \frac{{\eta_\phi}}{1+Q} \ll1 ~~,
\end{eqnarray}
where $\epsilon_\phi =({\Mpl^2}/{2})\left({V_{, \phi}}/{V}\right)^2$ and $\eta_\phi=\Mpl^2 ({V_{, \phi\phi}}/{V})$ are the usual cold inflationary (CI) slow-roll parameters. The quantity $N=\ln a$, with $a$  the scale factor, denotes the number of $e$-folds of inflation. Under the slow-roll approximation, the warm inflationary dynamics are then governed by the following:
\begin{eqnarray}
3H(1+Q)\dot{\phi} \approx V_{,\phi}~~, \\
4\rho_r \approx 3Q\dot{\phi}^2~~, \\
3H^2 \Mpl^2 \approx V~~,
\end{eqnarray}
where the radiation energy density can also be written in terms of the temperature $T$, since  $\rho_r= ({\pi^2}/{30}) g_{*} T^4$. 

For our studies of warm inflation \cite{Correa22}, we have adopted cubic dissipation $\Gamma= C T^3$ with $C$ a constant of dimension $\Mpl^{-2}$.  The dynamical equation  of  inflation with respect to the number of $e$-folds can be found in our earlier work \cite{Correa22}. 
\iffalse
The background equations, along with the definition of Q, can be combined to find a relation between $Q$ and $\phi$ ~\cite{Bastero-Gil:2019rsp}:
\begin{equation}
\label{PhiQrelation}
  \left( 4* 3^{2/c} \right) \left( \frac{ C_R}{C^{4/c}~ m_P^{~2+4/c}} \right ) ~ Q^{4/c-1} (1+Q)^2 =  \left ( \frac{V_{,\phi}^2}{V^{1+2/c}} \right )    
\end{equation}
The evolution of $\phi$, $Q$, and  $T/H$ w.r.t $N$ can be can be obtained using the slow-roll equations and finally for the cubic case one gets:
%%%%%%
\begin{equation}
    \frac{d({T}/{H}) }{dN} =\left( \frac{T}{H} \right) \frac{  \frac{(1-Q)}{4 Q}\frac{dQ}{dN} +\frac{3 \epsilon_{\phi} -\eta_{\phi} }{2}}{1+Q}
\end{equation}
with %%%%%%%%%
\begin{equation}
   \frac{dQ}{dN}=\frac{Q}{1+7Q} \left( 10 \epsilon_{\phi} -6\eta_{\phi} \right)
\end{equation}
so that%%%%%%%%%%%%
\begin{equation}
\frac{d(T/H)}{dN}= \frac{(T/H)}{1+7Q} ~ \left(\frac{8 Q+4}{1+Q} ~\epsilon_{\phi} -2 \eta_{\phi} \right) ~~.
\end{equation}
\fi

The power spectrum for curvature and tensor perturbations in the case of warm inflation with a cubic dissipation coefficient are respectively  given by  (see~\cite{Bartrum:2013fia} and refs. therein):
\begin{eqnarray}
       \cal{ P}_R&=&\left(\frac{H}{2\pi\dot{\phi}}\right)^2 \left( 1 +\frac{T}{H}F(Q) \right) {G(Q)} \, , \\
   {\cal{ P}}_T &= &2H^2/(\pi^2 \Mpl^2)\, , 
\end{eqnarray}
 with
\begin{equation}
    F(Q)
    \approx \frac{2\pi\sqrt{3}Q}{\sqrt{3+4 \pi Q}} ~~,
\end{equation}
and
\begin{equation}
    {G(Q)
    =  1 + 4.981 ~ Q^{1.946} + 0.127 ~ Q^{4.330} ~~, }
\label{GQ}
\end{equation}
where the $G(Q)$ approximates an exact numerical calculation of the effect of the coupling between the  inflating fluctuations and radiation \cite{Benetti17}.

\begin{table*}[ht]
\centering
\begin{tabular}{|l|l|
>{\columncolor[HTML]{FFFFFF}}l |
>{\columncolor[HTML]{FFFFFF}}l |l|l|}
\hline
$~~~Color~~~$                                     & $~~~N~~~$ & $~~~~\Lambda (m_p^4)~~~~$           & $~~~C_{\phi}~~~$ & $~~n_s~~$   & $~~r~~$                \\ \hline \hline
\textcolor{red}{Red}     & $~~~55~~~$  & $~~1.00\times 10^{-11}~~$   & $~~~50~~~$       & $~~0.964~~$ & $~~4.0\times10^{-3}~~$ \\ \hline \hline
\textcolor{blue}{Blue}   & $~~~63~~~$  & $~~7.87\times 10^{-12}~~$ & $~~~60~~~$       & $~~0.966~~$ & $~~4.5\times 10^{-4}~~$ \\ \hline \hline
\textcolor{green}{Green} & $~~~44~~~$  & $~~1.77\times 10^{-12}~~$ & $~~~40~~~$       & $~~0.966~~$ & $~~1.0\times 10^{-3}~~$ \\ \hline
\end{tabular}
\caption{(Color online) The inflationary observable for different sets of model parameter values. The different color codes are maintained in the plots. This is the same from \cite{Correa22}}. 
\label{tab1}
\end{table*}

Finally, the natural inflation potential  for the inflaton field is given by \cite{Freese}
\begin{equation}
    V(\phi)= \Lambda \left( 1+ \cos{\left(\frac{\phi}{f}\right)} \right)~,
\label{NIpot}
\end{equation}
where the inflaton $\phi$ is an "axion-like" field which is analogous to the Goldstone Boson of a  broken Peccei-Quinn-like
symmetry.  The parameter $f $ is the  symmetry breaking scale.   
This potential has been studied previously in the context of warm inflation  \cite{wnio,  wni1, wni2, Montefalcone:2022jfw, AlHallak:2022haa}. 
\iffalse
In this case, the primordial power spectrum of curvature fluctuations takes the form:
\begin{equation}
\begin{split}
P_{R}= &
\frac{
   \left[ C_{\phi}  \left(~1+ \cos{({\phi}/{f})}~\right) \right]^{4/3} 
   \left({\Lambda }/{m_{P}^4}\right)^{2/3}
   }{48*3^{2/3} ~ \pi ^2 ~ C_{r} ~Q^{1/3}}    
   \\& 
   \left(1+ \frac{3^{2/3}~ Q^{1/3}~ F(Q)}{
    \left[C_{\phi} (~1+ \cos{({\phi}/{f})}~) \right]^{1/3}
   \left({\Lambda }/{m_{P}^4}\right)^{1/6}}
   \right)
   G(Q)
\end{split}
\end{equation}
\fi 
In our study we have taken,  $C= {C_\phi}/{\Lambda^{1/2}}$ with $C_\phi$ being a dimensionless model parameter, and $\Lambda$ is the amplitude from Eq. (\ref{NIpot}). 
\begin{figure}
    \centering
    \includegraphics[height=3.0in,width=3.5in]{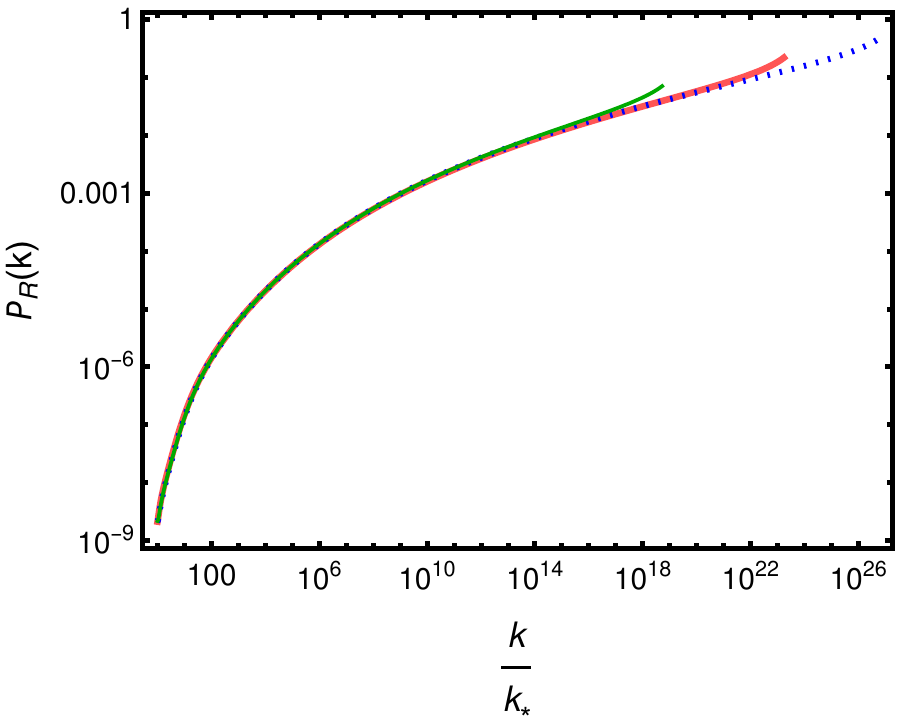}
    \caption{(Color online) Primordial power-spectra generated in the WNI the color code are same for the parameters as depicted in Table:\ref{tab1}.  }
    \label{fig:powerSpec}
\end{figure}

The primordial curvature power spectrum for three representative sets of parameters for this model, were given in \cite{Correa22} and are summarized here in Table 1.   The power spectra associated with these parameters  nicely produce PBHs  in the  desired  mass range whereby, a significant fraction (or all) of the current dark matter could be in the form of PBHs without violating observational constraints.

\section{Basics of Induced Gravity Waves}
\label{InducedGW}

%\subsection{Equations of motion}

To model the GWs a metric that includes both scalar and tensor perturbations is used. Within the Newtonian conformal gauge, the metric is then given  by \cite{Kohri:2018awv,Baumann:2007zm}:
\begin{eqnarray}
    ds^2 &=& a(\eta)^2\biggl[ -\left(1+2 \Phi\right)d\eta^2 \nonumber \\
    &+&\left(\left(1-2 \Psi\right)\delta_{ij}+ \frac{h_{ij}}{2}\right)dx^i dx^j
\biggr] ~~ ,
\end{eqnarray}
where $a$, is the scale factor, $\eta$ is the  conformal time, $\Phi $ and $\Psi$ are the scalar perturbations, while  $h_{ij}$ is the tensor perturbation added on top of  the  metric. The effect of the anisotropic stress  is neglected in the  following discussion since its contribution  is  small \cite{Baumann:2007zm}. Thus,  $\Phi=\Psi$.  

The action describing the tensor mode   is given by:
\begin{eqnarray}
    S=\frac{\Mpl^2}{32}\int d\eta~ d^3x~ a^2 \left[\left(h'_{ij}\right)^2-\left(\partial_l h_{ij}\right)^2\right]\, ,
    \label{acttion:graviton}
\end{eqnarray}
where the prime $'$ denotes differentiation w.r.t. conformal time $\eta$, while $\partial_l h_{ij}$ represents the derivative of $h_{ij}$ with respect to spatial coordinate $l$.

Next the tensor perturbation can be decomposed into its Fourier modes as:
\begin{eqnarray}
    h_{ij}(\eta, {\bf{x}})=\int \frac{d^3 ~k}{(2\pi)^{3/2}} \sum_{s=+, \times}\epsilon_{ij}^{s}(k)h^{s}_{{\bf k}}(\eta) e^{i{\bf k x}}~~,
\end{eqnarray}
where $\epsilon^{+}_{ij}(k)$ and $\epsilon^{\times}_{ij}(k)$ are  time-independent transverse traceless polarization  vectors, defined in an orthonormal  basis ($e_i({\bf k}), \bar{e}_i({\bf k})$) as 
\begin{eqnarray}
   \epsilon_{ij}^{+}(\bf k)&=&\frac{1}{\sqrt{2}}\left[ e_i({\bf k}) e_j({\bf k})-\bar{e}_i({\bf k}) \bar{e}_j({\bf k})\right] \\
   \epsilon_{ij}^{\times}(\bf k)&=&\frac{1}{\sqrt{2}}\left[ e_i({\bf k}) \bar{e}_j({\bf k})+\bar{e}_i({\bf k}) e_j({\bf k})\right]~~.
\end{eqnarray}
The dimensionless power spectrum for the tensor perturbation is then given by: 
\begin{eqnarray}
  \bigl\langle h_{\bf k}^\lambda (\eta) h_{\bf k'}^{\lambda'}(\eta)  \bigr\rangle = \delta_{\lambda \lambda'}\frac{2 \pi^2}{k^3} \delta(k+k') \mathcal{P}_{h}(k, \eta)\, ,
  \label{eq:Ten_PS}
\end{eqnarray}
where $\lambda, \lambda'= \{+, \times\}$.   
%%%%%%%%%%%%%%%%%%%%%%%%%%%%%%%%%%%%%%%%%%%%%%%%%%%%%%%%%%%%%%%%%%%%%%%%%%%%
What remains is to formulate equations of motion for the Fourier modes of the tensor  perturbations  $h_{\bf k}(\eta)$. The scalar perturbations in the gravitational potential $\Phi$ act as the source for the tensor equations of motion.  Hence, following \cite{Kohri:2018awv} we write:
\begin{align}
h''_{\bf k}(\eta) + 2 \mathcal{H} h'_{\bf k}(\eta)+ k^2 h_{\bf k}(\eta) =& 4 S_{\bf k}(\eta)~~,  \label{EOM_h} 
\end{align}
where the source term $S_{\bf k}$ is:
\begin{eqnarray}
S_{\bf k} &=& \int \frac{\text{d}^3 q}{(2 \pi)^{3/2}} e_{ij}({\bf k}) q_i q_j \biggl( 2\Phi_{\bf q}  \Phi_{{\bf k}-{\bf q}} +  \\
&& \frac{4}{3(1+w)} \left( \mathcal{H}^{-1} \Phi'_{\bf q} + \Phi_{\bf q}\right) \left( \mathcal{H}^{-1} \Phi'_{{\bf k}-{\bf q}} + \Phi_{{\bf k}-{\bf q}} \right)  \biggr) ~,\nonumber
\end{eqnarray}
The above has made use of: $-2 \dot{H} = \rho + P =( 1+w) \rho = 3 (1+w)H^2$, while $w = P/\rho$ is the usual equation-of-state parameter and $\mathcal{H}\equiv {a'}/{a}= a H $, is the Hubble parameter in conformal time. 
The Fourier modes of the gravitational potential $\Phi_{\bf k}$ are similar  to those of the tensor mode.
Next, a Green's function method can be used to solve for $h_{\bf k}(\eta)$,
\begin{align}
a(\eta) h_{\bf k}(\eta) = 4 \int^\eta \text{d}\bar{\eta} G_{\bf k}(\eta, \bar{\eta}) a(\bar{\eta}) S_{\bf k}(\bar{\eta})~~.
\label{eq:Ten_fluc}
\end{align}
Here,  the Green's function $G_{\bf k}(\eta, \bar{\eta})$ is the solution to
\begin{align}
G_{\bf k}''(\eta, \bar{\eta}) +\left( k^2 - \frac{ a''(\eta)}{a(\eta)}\right) G_{\bf k}(\eta, \bar{\eta}) = \delta (\eta - \bar{\eta}), \label{EOM_Green}
\end{align}
and derivatives are with respect to $\eta$.

The equation of motion  for the gravitational potential (e.g.~\cite{Mukhanov05}) is:
\begin{eqnarray}
&\Phi''_{\bf k}& + 3 \mathcal{H} (1 + c_{\text{s}}^2) \Phi'_{\bf k} + (2 \mathcal{H}'+(1+3 c_{\text{s}}^2)\mathcal{H}^2 +c_{\text{s}}^2 k^2) \Phi_{\bf k} 
\nonumber \\
&=& \frac{a^2}{2} \tau \delta S,  \label{EOM_Phi_complete}
\end{eqnarray}
where the sound speed $ c_{\text{s}}^2 = w$ and temperature  $\tau$ are defined via $\delta P = c_{\text{s}}^2 \delta \rho + \tau \delta S$, where  $S$ is the  entropy density.  
In the absence of entropy perturbations, the gravitational potential equation of motion reduces to
\begin{align}
\Phi''_{\bf k}(\eta) + \frac{6(1+w)}{(1+3w)\eta } \Phi'_{\bf k}(\eta) + w k^2 \Phi_{\bf k}(\eta)=0~~. \label{EOM_Phi}
\end{align}
The primordial value $\phi_{\bf k}$ is derived from the relation $\Phi_{\bf k} = \Phi(k \eta) \phi_{\bf k}$ where the transfer function $\Phi (k \eta)$ approaches unity well before the horizon entry.  The  primordial value  then relates to the curvature perturbation according to: 
\begin{align}
\langle \phi_{\bf k} \phi_{\bf k'} \rangle = \delta ({\bf k}+{\bf k}' ) \frac{2\pi^2}{k^3} \left( \frac{3+3w}{5+3w} \right)^2 \mathcal{P}_\zeta (k)~~,
\label{pzeta}
\end{align}
where the EoS parameter $w$ is evaluated before the horizon entry.
The ``primordial'' value $\phi_{\bf k}$ is also evaluated just before the horizon entry. 

The correlation function $\langle S_{\bf k} (\eta) S_{\bf k'}(\eta') \rangle$ is obtained by adopting Gaussian primordial curvature perturbations.  Finally, by comparing $\langle S_{\bf k} (\eta) S_{\bf k'}(\eta') \rangle$ with  $\mathcal{P}_h$, using Eqs. (\ref{eq:Ten_PS}) and (\ref{eq:Ten_fluc}) and doing some algebra, the power spectrum can be deduced from the curvature perturbation $P_\zeta$  \cite{Kohri:2018awv}:
\begin{eqnarray}
\mathcal{P}_h (\eta, k) &=&  4  
 \int_0^\infty \text{d}v \int_{\left| 1-v \right |}^{1+v}\text{d} u \left( \frac{4v^2 - (1+v^2-u^2)^2}{4vu} \right)^2 \nonumber \\
 &\times &I^2 (v,u,x) \mathcal{P}_\zeta ( k v ) \mathcal{P}_\zeta ( k u ), \label{P_h}
\end{eqnarray}
where $x\equiv k \eta$, and
\begin{align}
I(v,u,x)= \int_0^x \text{d}\bar{x} \frac{a (\bar{\eta})}{a(\eta)} k G_k (\eta, \bar{\eta}) f (v,u,\bar{x}),   \label{I}
\end{align}
with
\begin{eqnarray}
f (v ,u ,\bar{x}) &= & \frac{6(w+1)}{3w+5}\Phi(v\bar{x})\Phi(u\bar{x}) + \frac{6(1+3w)(w+1)}{(3w+5)^2} \nonumber \\
&\times &\left( \bar{x}\partial_{\bar{\eta}}\Phi(v\bar{x})\Phi(u\bar{x}) +\bar{x}\partial_{\bar{\eta}} \Phi(u\bar{x})\Phi(v\bar{x}) \right) \nonumber \\
& +& \frac{3(1+3w)^2(1+w)}{(3w+5)^2}\times \bar{x}^2 \partial_{\bar{\eta}}\Phi(v\bar{x})\partial_{\bar{\eta}}\Phi(u\bar{x}),\nonumber\\
\end{eqnarray}
and $\bar{x} \equiv  k \bar{\eta}$, while $\mathcal{H}=aH=2/[(1+3w)\eta]$. The function $f(u, v, \bar{x})$
contains the information about the source. 

A change of variables of  $u+v-1\rightarrow t$ and $u-v \rightarrow s$ recasts the integral \eqref{P_h} as 
\begin{eqnarray}
\mathcal{P}_h (\eta, k) &=&  2  
 \int_0^\infty \text{d}t \int_{ -1}^{1}\text{d} s \left( \frac{t(2+t)(s^2-1)}{(1-s+t)(1+s+t)} \right)^2 \nonumber \\
 &\times &I^2 (s,t,x) \mathcal{P}_\zeta ( k v ) \mathcal{P}_\zeta ( k u ) \label{P_h_st} ~~.
\end{eqnarray}

For GWs produced in a radiation-dominated universe in the late time limit ($ x\to\infty$), the oscillation average of $I^2(s,t,x)$  can be written as in \cite{Kohri:2018awv}, 
\begin{multline}
     \overline{I^2(s,t,x\to\infty)} = \frac{288(-5+s^2+t(2+t))^2}{x^2(1-s+t)^6(1+s+t)^6} \\
     \times \Biggl[\frac{\pi^2}{4}(-5+s^2+t(2+t))^2 \theta(t-(\sqrt{3}-1))  \\
      +\Bigl( ~-(t-s+1)(t+s+1)  \\
      +\frac{1}{2}(-5+s^2+t(2+t)) \log{\left|\frac{-2+t(2+t)}{3-s^2}\right|} ~\Bigl) ~\Biggl] ~~, \label{avg_I2}
\end{multline}
where $\theta(x)$ is the Heaviside step function. 

Combining Eqs.~\eqref{P_h_st} and \eqref{avg_I2} and further simplifying the integral by another change of variables ($t+1\rightarrow \sqrt{r}$), the following power spectrum is obtained: 
\begin{multline}
    \overline{\mathcal{P}_h(\eta, k)} =  \int_1^\infty \text{d}r \int_{ -1}^{1}\text{d} s \\
    \frac{72~(r-1)^2 \left(s^2-1\right)^2  \left(r+s^2-6\right)^2}{x^2 ~\sqrt{r} ~ \left(r-s^2\right)^8} \\
    \times \Biggl[  ~ \left(~ \left(r+s^2-6\right) ~\log \left(\left| \frac{3-r}{s^2-3}\right| \right)- 2 \left(r-s^2\right) ~ \right)^2 \\
    + \pi ^2 ~\left(r+s^2-6\right)^2 ~ \theta \left(\sqrt{r}-\sqrt{3}\right) ~ \Biggl] \\
    \times
    \mathcal{P}_\zeta\left(\frac{1}{2} k \left(\sqrt{r}-s\right)\right) 
    \mathcal{P}_\zeta\left(\frac{1}{2} k \left(\sqrt{r} + s\right)\right)~~.
\end{multline}

This can be numerically integrated using $\mathcal{P}_\zeta$ from Eq.~(\ref{pzeta}).

\section{Constraints on the GW energy density}
\label{GWcons}
 To compare WNI with  GW detector sensitivities, we need to calculate the present gravitational-wave closure contribution as a function of frequency. The power spectrum described in the previous section can be related directly to the energy density in gravitational waves. The GW energy density within the horizon is 
$\rho_{\text{GW}}(\eta)=\int \text{d}\ln k \rho_{\text{GW}}(\eta, k)$ and can be evaluated \cite{Maggiore00} as:
\begin{align}
\rho_{\text{GW}}=\frac{\Mpl^2}{16a^2} \left \langle \overline{h_{ij,k}h_{ij,k}} \right \rangle ,  \label{rho_GW}
\end{align}
 where the overline indicates an average over the oscillations. Considering the parity invariance for the polarization mode such that both have same contribution for the energy density. The fraction of the GW energy density  per logarithmic wave number,  $\Omega_{GW}(\eta,k)$ is then given by:
\begin{eqnarray}
\Omega_{\text{GW}}(\eta, k)&=& \frac{1}{\rho_{tot} (\eta)} \frac{d\rho_{\rm GW}(\eta, k) }{d\ln{k}}  \\
&= &\frac{\rho_{\text{GW}}(\eta,k)}{\rho_{\text{tot}}(\eta)}= \frac{1}{24} \left( \frac{k}{a(\eta)H(\eta)} \right)^2 \overline{\mathcal{P}_h(\eta, k)}~~, \nonumber
\label{Omega_GW}
\end{eqnarray}
where a sum has been made over the two polarization modes. During the radiation-dominated era, the term in parentheses simplifies to
\begin{equation}
\left( \frac{k}{a(\eta)H(\eta)} \right)^2= k^2\eta^2=x^2~~.
\end{equation}

To obtain the present spectrum of gravitational waves from Eq.~(\ref{Omega_GW}) following \cite{Kohri:2018awv} we deduce (see appendix~\ref{appendixA}):
\begin{align}
\Omega_{\text{GW}, 0}(k)= 0.39 \left( \frac{g_*(T_c)}{106.75} \right)^{-1/3} \Omega_{r,0} ~ \Omega_{\text{GW}}(\eta_c, k)~,
\label{Omega_GW_0}
\end{align}
where the subscript ($c$) denotes quantities evaluated when the perturbation is within the horizon during the radiation-dominated era when $\rho_{\rm GW}$ is  a constant fraction of radiation energy density. We use   $\Omega_{r,0} ~h^2 $ ($=4.18\times10^{-5}$) for the present closure contribution from photons and neutrinos.
Finally,  the wave-number $k$ is related to the frequency of gravitational waves by, 
\begin{align}
f=\frac{k}{2 \pi}=1.5\times10^{-15} \left( \frac{k}{1~\text{Mpc}^{-1}} \right) \text{Hz}~~.
\label{freq}
\end{align}

\begin{figure} 
%\centering
\includegraphics[height=3.1in,width=3.6in]{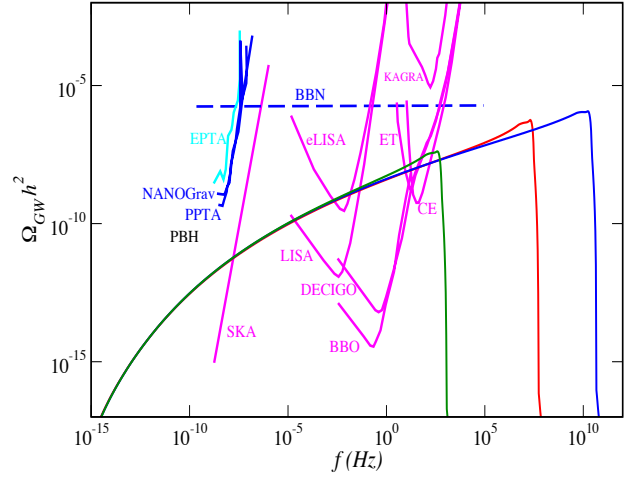}%
\caption{(Color online) Green, red and Blue solid lines show calculated closure contribution in gravitational radiation $\Omega_{GW}h^2$ from primordial black holes in WNI for $N = 44, 55, 60 $ $e$-folds of inflation.  These are  compared with various constraints as labeled.  Blue colors denote existing constraints from pulsar timing and BBN, while  pink colors show  constraints from future space-based and ground-based GW observatories.}
\label{fig:omega}
\end{figure}

Fig.~\ref{fig:omega} shows the present gravitational closure contribution as a function of frequency.   The continuous green, red and blue lines show calculated contribution in gravitational radiation $\Omega_{GW}h^2$ from primordial black holes for $N = 44, 55, 63$ $e$-folds of inflation respectively.  Note the sharp drop-off in the power once the scale of PBH formation is obtained.

   Various lines on Fig.~\ref{fig:omega} indicate current (blue lines) and anticipated future (pink lines) constraints from various GW observatories as labeled.    The jagged blue lines indicate existing pulsar timing array constraints from EPTA~\cite{EPTA}, NANOGrav~\cite{NANOGrav}, and PPTA~\cite{PPTA}.  The horizontal blue dashed  line shows the BBN upper bound on the energy density in gravitational waves, $\Omega_{\text{GW}}h^2 < 1.8 \times 10^{-6}$ (95\% C.L.) deduced in \cite{Kohri:2018awv}.
    
    The pink lines  show sensitivity curves~\cite{GW-review} of various future GW observations reproduced from Ref.~\cite{Moore14}.  The curves are from SKA~\cite{SKA}, eLISA~\cite{eLISA}, LISA~\cite{LISA}, BBO~\cite{BBO}, DECIGO~\cite{DECIGO}, Einstein Telescope~\cite{ET}, Cosmic Explorer~\cite{CE}, and KAGRA~\cite{KAGRA}.

As seen from this figure, the predicted contribution from PBH GWs easily satisfies current constraints from pulsar timing and BBN.  Perhaps, more interesting is the fact that  pending space-based detectors such as LISA, BBO, and DECIGO will have sufficient sensitivity to detect this contribution from PBH gravity waves.  Even next-generation ground-based detectors like the Einstein Telescope and the Compton Explorer may get a glimpse of this possible GW background.

%\vfill\eject
\section{Conclusion:}
\label{conclusion}
We have calculated the contribution to the closure density $\Omega_{\rm GW}h^2$ from the energy density in gravitational waves for models of warm natural inflation that produce PBHs in a mass range that could account for as much as all of the presently inferred dark matter. The case of PBHs in cold inflation has been studied extensively in the  recent literature such as: \cite{Sasaki:2018dmp, Inomata:2017okj, Mishra:2019pzq, Bhattacharya:2021wnk, 
cipbh2}. However, we have shown in particular that the contribution of $\Omega_{\rm GW}$ from WNI satisfies all existing constraints from BBN and pulsar timing.  Moreover, we show the interesting result that this cosmic background in GWs may be detectable in the next generation of space-based and ground-based  gravitational wave interferometers.

 The observation of GWs predicted by this model could point towards indirect evidence of the warm inflationary paradigm. Of course, one should also explore many avenues to check the shape of the secondary GWs produced due to the enhancement in the primary scalar power spectrum. For example, one interesting aspect that one can check is the effect on the GW spectrum in the case of resonant particle production during inflation as described in \cite{Mathews:2015daa, Gangopadhyay:2017vqi}. 
 
 Also, a more general form of pseudo-Nambu Goldstone Boson(pNGB) inflaton has been studied in \cite{Croon:2015fza} and studied in alternative scenarios in \cite{Bhattacharya:2018xlw, Khan:2022odn, Gangopadhyay:2022vgh, Gangopadhyay:2022dbm}.  This has been  dubbed as Goldstone inflation. In this case, Natural inflation is just a limiting case of the more general Goldstone inflation. It will be of interest to study this more general model in the context of the WI and study further the PBH production and GW production associated with it. 
 
 Furthermore, a reconstruction of the inflationary potential in the WI paradigm while keeping the PBH production in mind to account for the total DM density could lead to interesting results as in the case of \cite{Bhattacharya:2019ryo}. 
 
 Finally, we note a recent suggestion that the production of PBHs faces a no-go theorem in the case of  single-field cold inflation \cite{Choudhury:2023vuj}.  We emphasize, however,  that  the production of PBHs and consequently GWs in the context of WNI as discussed here  is both allowed and quite inevitable. Thus, testing this theory following the path of \cite{Choudhury:2023vuj, Choudhury:2023jlt, Choudhury:2023rks, Choudhury:2023hvf, Kristiano:2022maq, Kristiano:2023scm} (though there are counter-arguments presented in \cite{Riotto:2023hoz}) could lead to  interesting insight into the physics of the early inflationary universe. The authors plan to consider these in  future work.
%%%%%%
\vspace{0.3cm}

%%%%%%%
\textit{Acknowledgments.}---
The authors would like to thank M. Sami, and Yogesh for the useful discussions. N.J. is thankful to Sirshendu for his help.  Work at the University of Notre Dame supported by DOE nuclear theory grant DE-FG02-95-ER40934. Work of M.R.G. is supported
by DST, Government of India under the Grant Agreement number IF18-PH-228 and by Science and Engineering Research Board (SERB), DST, Government of India under the Grant Agreement number CRG/2022/004120 (Core Research Grant). N.J. is supported by the National
Postdoctoral Fellowship of the Science and Engineering Research Board (SERB), Department of Science and Technology (DST), Government of India, File No. PDF/2021/004114. 
\appendix
\section{}
\label{appendixA}
Here we show the derivation of $\Omega_{GW}$. 
  Following \cite{Kohri:2018awv} we write. 
 \begin{eqnarray}
   \Omega_{\rm GW,0}(k) &=& \frac{\rho_{\rm GW}(\eta_0,k)}{\rho_{\rm tot}(\eta_0)}= \frac{\rho_{\rm GW}(\eta_0, k)}{\rho_{\rm GW}(\eta_c, k)} \frac{\rho_{\rm GW}(\eta_c, k)}{\rho_{\rm tot} (\eta_0)}\nonumber\\ 
   &=&  \frac{\rho_{\rm GW}(\eta_0, k)}{\rho_{\rm GW}(\eta_c, k)} \frac{\rho_{\rm GW}(\eta_c, k)}{\rho_{\rm tot} (\eta_c)}\frac{\rho_{\rm tot}(\eta_c)}{\rho_{\rm tot}(\eta_0)}~.
\end{eqnarray}
During the radiation dominated era, $\rho_{\rm tot}(\eta_c) \approx \rho_r(\eta_c)$ and also $\rho_{\rm GW} \sim a^{-4}$. Thus, can we write
\begin{eqnarray}
    &=& \frac{a(\eta_c)^4}{a(\eta_0)^4}\Omega_{\rm GW}(\eta_c, k) \frac{\rho_r (\eta_c)}{\rho_{\rm tot}(\eta_0)}\\
    &=& \frac{a(\eta_c)^4}{a(\eta_0)^4}\Omega_{\rm GW}(\eta_c, k) \frac{\rho_r (\eta_0)}{\rho_{\rm tot}(\eta_0)}\frac{\rho_{r }(\eta_c)}{\rho_r (\eta_0)}\\
    &=&\frac{a(\eta_c)^4}{a(\eta_0)^4}\Omega_{\rm GW}(\eta_c, k) \Omega_{r, 0}\frac{\rho_{r }(\eta_c)}{\rho_r (\eta_0)}~~.
    \label{final}
\end{eqnarray}
Now, from the conservation of entropy we get
\begin{eqnarray}
    \frac{\rho_r (\eta_c)}{\rho_r (\eta_0)} &=&\frac{g_{*,c}}{g_{*,0}} \left(\frac{T_c}{T_0}\right)^4\\
   &=&\frac{g_{*,c}}{g_{*,0}}  \left(\frac{g_{*S,0}}{g_{*S,c}}\right)^{4/3}\frac{a(\eta_0)^4}{a(\eta_c)^4}~~,
\end{eqnarray}
where $g_{*, c}$ is the number of relativistic degrees freedom at temperature $T_c$, similarly $g_{*S}$ is the same for entropy density. Before $e^+-e^-$ pair annihilation during the radiation era, $ g_{*S,c} \approx g_{*,c} $, Thus, we deduce: 
\begin{eqnarray}
   \Omega_{\rm GW,0}(k)
   &=&\left( \frac{g_{*,c}}{g_{*,0}}\right)
  \biggl( \frac{(g_{*S,0}}{g_{*S,c}}\biggr)^{4/3}
   \Omega_{r,0} ~ \Omega_{\text{GW}} \nonumber\\
  &=& 0.39 \left(\frac{g_{*,c}}{106.75}\right)^{-1/3} \Omega_{r,0} ~ \Omega_{\text{GW}}(\eta_c, k)~,
\end{eqnarray}
where, the latter equation makes use of  standard values $g_{*,0}= 3.36$ and $g_{*S, 0}= 3.91$, and 106.75 is the number of degrees of freedom from standard-model particles. %Finally using  Eq. \ref{final} we obtain
%\begin{eqnarray}
% \Omega_{\rm GW,0}(k) = 0.39 \left(\frac{g_{*,c}}{106.75}\right)^{-\frac{1}{3}} \Omega_{r,0} ~ \Omega_{\text{GW}}(\eta_c, k)\, .
% \end{eqnarray}
%\bibliographystyle{utphys}
%\bibliography{wipbh}

%%%%%%%%%%
%%%%%%%%%%%%

\clearpage
\appendix
%\onecolumngrid
%\section*{Supplemental Material}

%%%%%%%%%%%%%%%%%%%%%%%%%%%%%%%%%%%%%%%%%%%%%%%%%%%%%%%%%%%%%%%%%%%%%%%%%%%%%%%
\end{document}